\documentstyle{espart}
\begin{document}

\def\lsim{\thinspace{\hbox to 8pt{\raise
-5pt\hbox{$\sim$}\hss{$<$}}}\thinspace}
\def\rsim{\thinspace{\hbox to 8pt{\raise
-5pt\hbox{$\sim$}\hss{$>$}}}\thinspace}


\begin{frontmatter}

\title {Mean Field Studies of Exotic Nuclei}

\author{ C. R. Chinn}
\address {Department of Physics and Astronomy
Vanderbilt University, Nashville, TN  37235 }

\author{ A. S. Umar}
\address {Department of Physics and Astronomy
Vanderbilt University, Nashville, TN  37235
and Center for Computationally Intensive Physics,
Oak Ridge National Laboratory, Oak Ridge, TN  37831-6373}

\author{M.~Valli\`eres,}
\address {Department of Physics and Atmospheric Sciences,
          Drexel University, Philadelphia, PA 19104}

\author{M. R. Strayer}
\address {Physics Division, Oak Ridge National Laboratory
Oak Ridge, Tennessee  37831  and
Center for Computationally Intensive Physics,
Oak Ridge National Laboratory, Oak Ridge, TN  37831-6373}

\begin{abstract}
{Full three dimensional static and dynamic mean field calculations
using collocation basis splines with a Skyrme type Hamiltonian are
described.  This program is developed to address the difficult
theoretical challenges offered by exotic nuclei.  Ground state and
deformation properties are calculated using static Hartree-Fock,
Hartree-Fock+BCS and constrained Hartree-Fock models.  Collective
properties, such as reaction rates and resonances, are described
using a new alternate method for
evaluating linear response theory, which is constructed directly
on top of the static calculation.  This provides a consistent
description of the ground state, deformation and collective
nuclear properties.  Sample results are presented for the giant multiple
resonances of $^{16}$O.  }
\end{abstract}

\end{frontmatter}


\section{Introduction}

With the recent and future advent of new experimental facilities,
such as the Radioactive Ion Beams facility at ORNL and GammaSphere,
exploratory detailed studies of many exotic nuclei will need
correspondingly accurate theoretical investigations.  Examples of
some of the properties of nuclei which will be studied include
effective interactions, masses and ground state properties,
multi-dimensional energy surfaces and reaction rates.  The results
of these studies may also impact and address many astrophysical
questions.

The calculations discussed in this paper can be separated into two
basic categories.  These are static mean field theories, such as
Hartree-Fock, Hartree-Fock-Bogolyubov and constrained Hartree-Fock
calculations; and using dynamic linear response theory to study
large-scale collective motion and reaction rates.

Because of new capabilities in computer technology as well as
innovations in numerical algorithms, a much greater level of sophistication
can be incorporated into numerical studies than in the past.  In
the studies presented in this paper, we incorporate the following
ingredients.  Collocation lattice representations using basis
splines are used instead of traditional grid representations or
basis expansions.  A full unrestricted three-dimensional calculation
is performed, where no assumptions about spatial symmetries are
needed.  Spin degrees are included and one need not impose time
reversal symmetry.  As a result, to study nuclei with a large number
of single particle states one requires grand challenge level
computing resources.

Collocation basis splines provide a superior representation than
traditional grid techniques.  The reason for this is that with
basis splines, the gradient operator is represented as an
operation upon a set of basis functions, which is much more
accurate than the finite difference methods used in traditional
calculations.  At the same time collocation basis splines allow
for the advantageous use of lattice representations.  Here a full
three-dimensional collocation lattice is used.

It may be argued that there are existing accurate mean field
calcualtions which use alternate basis expansions, such as sets of
harmonic oscillator functions.  One finds though that although
these methods provide for accurate studies of more traditional nuclei,
when one addresses exotic nuclei near the drip lines, such
calculations begin to fail to accurately represent the physics.
This can be traced to the fact that exotic nuclei, because of
small binding energies, have very extended density distributions.
One finds the harmonic oscillators bases typically used, due
to their natural locality,
are less able to represent accurately such extended objects than
can a grid representation \cite{li11}.

An alternate method has been developed to solve the linear response
theory, which is a refinement of the technique used in ref.~\cite{bf}.
Instead of solving a specific set of equations, the
response of the system is derived using time-dependent
Hartree-Fock [TDHF] theory.  In this case a specific
time-dependent perturbing piece is added to the static hamiltonian
to give $H_{tot}(t)$.  The static Hartree-Fock solution is then
time-evolved using $H_{tot}(t)$ in a TDHF calculation.  The
Fourier transform of the result gives the response of the system
at all energies.  In this scheme one recovers both the response
spectrum as well as the total transition probability amplitude
corresponding to a given specific collective mode.

The main advantage of this approach is that the dynamical linear
response calculation is constructed directly on top of the static
Hartree-Fock calculation and hence the static and dynamical
calculations are calculated using the same hamiltonian
description and the same degree of complexity.
Therefore, there is a complete consistency between
the static ground state of the system and the response
calculations.  One can then provide a coherent description of
static properties of nuclei and of dynamical properties.  This is
important for example in $\beta-$decay calculations of exotic
nuclei, where reliable predictions are very sensitive to the
deformation properties of the nucleus.  Hence in this formalism
consistent predictions of both the deformation and reaction rate
properties are possible.

Pairing correlations can be included in both the static and
dynamic formalisms.  Presently, the BCS and Lipkin-Nogami
prescriptions for including pairing have been used.  Eventually we
would like to use a self-consistent description of pairing
correlations, which would enable us to perform quasi-RPA studies
of $\beta-$decay for nuclei near the drip lines.  With the
recent advent of radioactive beam facilities, one would like to
provide accurate and physical predictions of properties of exotic
nuclei.

Section 2 contains the theoretical discussion.
Results are given in Section~3 followed by a summary
and conclusion in Section~4.


\section{Theoretical Discussion}

\subsection{Continuous Equations}

The details of the derivation of the Hartree-Fock equations can be found
in \cite{ASU,DaKo,HN,KDM}. The result for a many-body Hamiltonian containing a
one-body kinetic energy and two- and three-body momentum dependent
potential terms is a coupled set of non-linear partial differential eigenvalue
equations,
\begin{equation}
{\bf h}{\bf \chi}_{\alpha}=
     \epsilon_{\alpha}{\bf \chi}_{\alpha}\;,~~~~~~
  {\bf \chi}_{\alpha}= \left( \begin{array}{c}
                    \chi_{\alpha}^+ \\
                   \chi_{\alpha}^- \end{array} \right) \;,
\label{hfeq}
\end{equation}
where ${\bf \chi}_{\alpha}$ is a two-component vector (spinor)
The effective nucleon-nucleon interaction is chosen to be of the
zero-range Skyrme type, hence the Hamiltonian ${\bf h}$ can
be written in the following form (using natural units
$\hbar=1$, $c=1$, $m=1$)
\begin{eqnarray}
{\bf h}&=&-\frac{1}{2}{\pol \nabla}^2+
       W(\rho,\tau,{\pol j},{\pol J})\;, \nonumber \\
W&=&V_N({\pol r})+V_C({\pol r})\;,
\end{eqnarray}
where $V_N$ is the nuclear potential depending on various currents and
densities and $V_C$ is the Coulomb interaction.
The densities and currents depend on the states $\chi_{\alpha}$
and are explicitly given by
\begin{eqnarray}
\rho({\pol r})&=&\sum_{\alpha} w_\alpha\{
     |\chi_{\alpha}^+({\pol r})|^2 +
          |\chi_{\alpha}^-({\pol r})|^2 \} \\
 &~ & \nonumber \\
\tau({\pol r})&=&\sum_{\alpha} w_{\alpha}\{
                     |\nabla\chi_{\alpha}^+({\pol r})|^2 +
                     |\nabla\chi_{\alpha}^-({\pol r})|^2 \}
\\
& & \nonumber \\
{\pol j}({\pol r})&=&\sum_{\alpha} w_{\alpha}\{Im[
 {\chi^+_{\alpha}}^*({\pol r})
     {\pol \nabla}{\chi^+_{\alpha}}({\pol r})+
 {\chi^-_{\alpha}}^*({\pol r}){\pol \nabla}
       {\chi^-_{\alpha}}({\pol r}) \}
\\
& & \nonumber \\
{\pol J}({\pol r})&=&-i\sum\limits_{\alpha \atop \mu\mu'=\pm}
                w_{\alpha} {\chi_{\alpha}^{\mu}}^*({\pol r})
          ({\pol \nabla} \times {\pol \sigma})
            {\chi_{\alpha}^{\mu'}}({\pol r})\;.
\label{densit}
\end{eqnarray}
$V_C$ requires the solution of the Poisson equation in
three-dimensional geometry,
\begin{equation}
\nabla^2V_C({\pol r})=-4\pi e^2\rho({\pol r})\;.
\label{coul}
\end{equation}

As can be seen from above, the solution of the system of equations
(\ref{hfeq}) has to be obtained self-consistently and an accurate solution
requires a good representation of various derivatives of the states
$\chi_{\alpha}$.
Currently, most HF and TDHF calculations are performed using
finite difference lattice techniques.
It is desirable to investigate
higher-order interpolation methods which result in the improvement
of the overall accuracy and reduction in the total number of lattice
points.
The lattice solution of differential equations on a discretized
mesh of independent variables may be viewed to proceed in two
steps: (1) Obtain a discrete representation of the functions and
operators on the lattice. (2) Solve the resulting lattice equations
using iterative techniques.
Step (1) is
an interpolation problem for which we could take advantage of the
techniques developed using the spline functions \cite{Boor,bottcher,umar3}.
The use of the
spline collocation method leads
to a matrix-vector representation on the collocation lattice with
a metric describing the transformation properties of the collocation
lattice.

\subsection{Solution of the Discrete HF Equations}

The solution of the HF equations (\ref{hfeq}) is found by
using the damped relaxation method described in Refs. \cite{BSU,umar1}
\begin{equation}
\chi_{\alpha}^{k+1}={\cal O}\left[ \chi_{\alpha}^k-
x_0 {\bf D}(E_0)\left({\bf h}^k-
\epsilon_{\alpha}^k\right)\chi_{\alpha}^k \right]\;,
\label{numhf}
\end{equation}
where ${\cal O}$ stands for Gram-Schmidt orthonormalization.
The preconditioning operator ${\bf D}$ is chosen to be \cite{BSU,umar1}
\begin{eqnarray*}
{\bf D}(E_0)&=&\left[ 1+\frac{{\bf T}}{E_0}\right]^{-1} \\
&\approx&
\left[ 1+\frac{{\bf T}_x}{E_0}\right]^{-1}
\left[ 1+\frac{{\bf T}_y}{E_0}\right]^{-1}
\left[ 1+\frac{{\bf T}_z}{E_0}\right]^{-1}\;,
\end{eqnarray*}
where ${\bf T}$ denotes the kinetic energy operator.
The solution is obtained through an iterative scheme as outlined below:

\begin{itemize}
\begin{enumerate}
\item Guess a set of orthogonal single-particle states
\item Compute the densities (\ref{densit})
\item Compute the Hartree-Fock potential
\item Solve the Poisson equation
\item Perform a damping step (\ref{numhf}) without orthogonalization
\item Do a Gram-Schmidt orthogonalization of all states
\item Repeat beginning at step (ii) until convergence
\end{enumerate}
\end{itemize}

In practical calculations we have used
the damping scale value $x_0\approx 0.05$ and the energy cutoff
$E_0\approx 20.0$.
As a
convergence criteria we have required the fluctuations in energy
\begin{equation}
\Delta E^2\equiv\sqrt{\langle H^2\rangle-\langle H\rangle^2}\;
 \label{eq:var}
\end{equation}
to be less than $10^{-5}$. This is a more stringent condition than
the simple energy difference between two iterations, which is about
$10^{-10}$ when the fluctuation accuracy is satisfied.
The calculation of the HF Hamiltonian also requires the
evaluation of the Coulomb contribution given
by Eq. (\ref{coul}).
Details of solving the Poisson equations using the
splines are given in Ref. \cite{bottcher,umar3}.

\subsection{Collocation Basis Splines}

The static and time-dependent Hartree-Fock calculations are
performed using a collocation spline basis in a three-dimensional
lattice configuration.  Basis splines allow for the use of a
lattice grid representation of the nucleus, which is much easier
to use than alternate basis techniques, such as multi-dimensional
harmonic oscillators.  Also, for studies of exotic nuclei, because of
weak binding, the density distributions tend to extend to large
distances and hence one finds a large sensitivity to a harmonic
oscillator basis, while for a lattice grid representation one
needs to simply increase the size of the box.  Traditional grid
representations typically use finite difference techniques to
represent the gradient operator.  Collocation basis splines allow
for the gradient operator to be represented by its action upon a
basis function in a matrix form.  This collocation method gives a
much more accurate representation of the gradient, while
maintaining the convenience of a lattice grid and hence
provides a much more accurate calculation in the end.

An $M$th order spline function denoted by $B_i^M$ is constructed
from piecewise continuous polynomials up to order $M-1$.  The set
of points or {\it knots} $\{x_i\}$ consists of the points where
the spline functions are joined continuously up to the $(M-2)$
derivative.  The basis spline functions have minimal support in
that the $i$th spline functions is nonzero only in the interval
$(x_i,x_{i+M})$, where the spline function, $B_i^M$, is labelled
by the first knot.  For the case of using static boundary
conditions for the space containing the $N+1$ knots in
one dimension, there must be $M$ nonzero spline functions in each
interval, hence $N+2M-1$ total spline functions make up the full
basis, where $M-1$ spline functions extend beyond each boundary.
One can also impose periodic boundary conditions, where there will
then be $N$ total spline functions.

A function, $f(x)$ continuous in the interval $(x_{min},x_{max})$
is expanded in terms of the spline basis functions:
\begin{equation}
f(x) = \sum_i^{N+2M-1} B_i^M(x) c^i .  \label{eq:1.1}
\end{equation}
The expansion coefficients, $c^i$ are derived from $f(x)$ by
evaluating $f(x)$ at a specific set of points called {\it
collocation points}, $\{x_{\alpha}'\}$.  There are various ways
of choosing the $\{x_{\alpha}'\}$.  For odd order splines we have
chosen the collocation points to lie at the center of each knot
interval within the range $(x_{min},x_{max})$.
\begin{eqnarray*}
x_{min} & = & x_M ,  \\
x_{\alpha}' & = & \frac{x_{i+M-1} + x_{i+M}}{2}, ~~i = \alpha,
       ~~~\forall \alpha = 1,...,N
\end{eqnarray*}
By evaluating eq.~(\ref{eq:1.1}) at $\{x_{\alpha}'\}$, a set of
linear equations are constructed which constrain the
coefficients, $c^i$:
\begin{equation}
f(x_{\alpha}') = \sum_i^{N+2M-1} B_i^M(x_{\alpha}') c^i .
\label{eq:1.2}
\end{equation}
Since there are $N+2M-1$ unknown coefficients in the case of
static boundary conditions and $N$ points
$x_{\alpha}'$, it is necessary to introduce $2M-1$ additional
constraining equations as boundary conditions.
Combining the functions
$B_i^M(x_{\alpha}')$ and the boundary conditions into a square
invertible matrix, ${\bf B}$, the coefficients, $c^i$ can be
expressed as:
\begin{equation}
c^i = \sum_{\alpha} \left[ {\bf B}^{-1}\right] ^{i\alpha}
                          f_{\alpha} , \label{eq:1.3}
\end{equation}
where $f_{\alpha} \equiv f(x_{\alpha}')$ is the collocation
representation of the function, $f(x)$.

Consider the action of an operator upon a function:
\begin{equation}
{\widehat O} f(x) = \sum_i^{N+2M-1} \left[ {\widehat O}
                     B_i^M(x) \right] c^i . \label{eq:1.4}
\end{equation}
If we evaluate the above expression at the collocation points and
substitute in eq.~(\ref{eq:1.3}) for the $c^i$ the following is
obtained.
\begin{equation}
{\widehat O} f(x_{\alpha}') = {\widehat O} f_{\alpha} =
         \sum_i^{N+2M-1} \left[
         {\widehat O} B_i^M(x_{\alpha}') \right]
         \sum_{\beta} \left[ {\bf B}^{-1}\right]^{i\beta} =
         \sum_{\beta}
       {\widehat O}_{\alpha}^{\beta} \, f_{\beta} , \label{eq:1.5}
\end{equation}
where now the quantity, ${\widehat O}_{\alpha}^{\beta}$ is the
collocation representation of the operator, ${\widehat O}$ on
the lattice.
\begin{equation}
{\widehat O}_{\alpha}^{\beta} \equiv \sum_i^{N+2M-1} \left[
         {\widehat O} B_i^M(x_{\alpha}') \right]
          \left[ {\bf B}^{-1}\right]^{i\beta}
\end{equation}
Note that the representation, ${\widehat O}_{\alpha}^{\beta}$,
is not a sparse matrix.

The function, $f(x)$, and the operator, ${\widehat O}$, can both
be represented on a lattice, {\it i.e.} the collocation points,
through the use of the collocation basis spline method.  This
holds true for gradient operators, where the gradient of the
basis spline functions is required.  The basis spline functions,
$B_i^M(x)$ and their derivatives,
$\frac{\partial^n B_i^M(x)}{\partial x^n}$ can be evaluated at
the collocation points using iterative techniques.  Through
similar methods one can obtain the appropriate integration
weights:
\begin{equation}
 I  = \int_a^b f(x) dx
 = \sum_\alpha w^\alpha f(x_\alpha),  \nonumber
\end{equation}
where the weights are given as :
\begin{equation}
 w^{\alpha} \equiv \sum_i^{N+2M-1} \int_a^b
         B_i^M(x) dx
          \left[ {\bf B}^{-1}\right]^{i\alpha}.
\end{equation}
For more details on the collocation basis spline method
please see Ref.~\cite{umar3}.

\subsection{Linear Response Theory}

The linear response equations can be derived from a specific
functional time-dependent perturbation of the
TDHF equations \cite{fetter}.
To begin the proof a solution to the static Schr{\"o}dinger
equation is written as follows:
\begin{equation}
{\widehat H}|\psi_s(0)\rangle = E |\psi_s(0)\rangle . \label{eq:3.1}
\end{equation}
A time-dependent perturbing function is added to the
static hamiltonian:
\begin{equation}
{\widehat H}_{tot} = {\widehat H} + {\widehat H}_{ex}(t) .
\label{eq:3.2}
\end{equation}
The external piece is defined as:
\begin{eqnarray}
{\widehat H}_{ex}(t) & = & {\widehat F} f(t) \nonumber \\
  & = & \left[ \int d^3x\, {\widehat n}(x,t)\, F(x)\right]
  f(t) , \label{eq:3.3}
\end{eqnarray}
where ${\widehat n}(x,t)$ is the number density operator.
$F(x)$ corresponds to some particular collective mode.
The functions, $F(x)$ and $f(t)$ will be chosen later.

At some time $t=t_0$ the external piece of the Hamiltonian is
turned on where $|{\bar \psi}_s(t)\rangle$ is the solution
to the time-dependent Schr\"{o}dinger equation:
\begin{equation}
\imath\hbar\frac{\partial}{\partial t} |{\bar \psi}_s(t)\rangle
 = \left[ {\widehat H} + {\widehat H}_{ex}(t)\right]
     |{\bar \psi}_s(t)\rangle .  \label{eq:3.4}
\end{equation}
Here the subscript `$s$' refers to the Schr\"{o}dinger picture.
The solution for the state vector can be written in the
following iterative fashion.
\begin{equation}
|{\bar \psi}_s(t)\rangle = e^{-\imath{\widehat H}t/\hbar}
 |\psi_s(0)\rangle - \frac{\imath}{\hbar} e^{-\imath{\widehat H}t/\hbar}
 \int_{t_0}^{t} dt'{\widehat H}_{ex}^I(t')|\psi_s(0)\rangle
  + ... ~,\label{eq:3.8}
\end{equation}
where the superscript `$I$' refers to the interaction picture.
The expectation value of any operator, ${\widehat O}(t)$, is
equal to
\begin{eqnarray}
\langle{\bar \psi}_s(t)|{\widehat O}_S(t)|{\bar \psi}_s(t)\rangle
 & = & \langle{\bar \psi}_s(0)|{\widehat O}_S(0)
|{\bar \psi}_s(0)\rangle +  \\
 & ~ & + \langle{\bar \psi}_s(0)|\frac{\imath}{\hbar}
 \int_{t_0}^{t} dt'\left[{\widehat H}_{ex}^I(t'),
{\widehat O}_I(t)\right] |{\bar \psi}_s(0)\rangle + ...~.
      \label{eq:3.9} \nonumber
\end{eqnarray}
The linear approximation is made such that terms beyond first
order in ${\widehat H}_{ex}^I$ are neglected.

If we define the operator ${\widehat O}(t)$ to be the number
density operator, then using eq.~(\ref{eq:3.3}) the fluctuation
in the density becomes:
\begin{eqnarray}
\delta\langle {\widehat n}(x,t)\rangle & = &
\langle{\bar \psi}_s(t)|{\widehat O}_S(t)|{\bar \psi}_s(t)\rangle
- \langle{\bar \psi}_s(0)|{\widehat O}_S(0)|{\bar \psi}_s(0)\rangle
   \nonumber \\
 & = & \langle{\bar \psi}_s(0)|\frac{\imath}{\hbar}
\int_{t_0}^{t} dt' \int d^3x' F(x') f(t')
\left[ {\widehat n}_I(x',t'),{\widehat n}_I(x,t)\right]
 |{\bar \psi}_s(0)\rangle  \label{eq:3.10}
\end{eqnarray}

The retarded density correlation function is defined as:
\begin{equation}
\imath D^R(x,t;x',t') = \theta(t-t') \frac{\langle\psi_0 |
 \left[ {\widetilde n}_H(k), {\widetilde n}_H(k') \right]
  | \psi_0\rangle}{\langle\psi_0 |\psi_0\rangle} , \label{eq:3.11}
\end{equation}
where ${\widetilde n}_H={\widehat n}_H-\langle{\widehat n}_H\rangle$
is the deviation of the number operator in the Heisenberg
picture.  The density fluctuation can be written as:
\begin{equation}
\delta\langle {\widehat n}(x,t)\rangle = \frac{1}{\hbar}
         \int_{-\infty}^{\infty}
dt' \int d^3x' D^R(x,t;x',t') F(x') f(t') . \label{eq:3.12}
\end{equation}
Using the Fourier representation of $\theta(t-t')$, the Fourier
transform of the density correlation function is:
\begin{eqnarray}
\imath D^R(x,x';\omega) & = & \int_{-\infty}^{\infty} d(t-t')
 e^{\imath\omega(t-t')} \imath D^R(x,t;x',t') \nonumber \\
 & = & \sum_n \left\{
\frac{\langle\psi_0| {\widetilde n}_S({\pol x}|\psi_n\rangle
\langle\psi_n| {\widetilde n}_S({\pol x'}|\psi_0\rangle }
{\omega - \frac{E_n-E_0}{\hbar} + \imath\eta} -
\frac{\langle\psi_0| {\widetilde n}_S({\pol x'}|\psi_n\rangle
\langle\psi_n| {\widetilde n}_S({\pol x}|\psi_0\rangle }
{\omega + \frac{E_n-E_0}{\hbar} + \imath\eta} \right\} ,
  \label{eq:3.13}
\end{eqnarray}
where $|\psi_n\rangle$ represents the full spectrum of excited
many-body states of $\widehat H$.

The Fourier transform of the density fluctuation then becomes:
\begin{eqnarray}
\delta\langle {\widehat n}(x,\omega)\rangle & = &
     \int_{-\infty}^{\infty} dt
  e^{\imath\omega t} \delta\langle n(x,t)\rangle \nonumber \\
 & = & \frac{1}{\hbar} \int d^3x' D^R({\pol x},{\pol x'};\omega)
F(x') f(\omega) , \label{eq:3.14} \\
{\rm where}~~~~~~~~~~ f(\omega) & = & \int_{-\infty}^{\infty} dt'
       e^{\imath\omega t'} f(t') \nonumber
\end{eqnarray}

The linear response structure function, $S(\omega)$ is derived to
be:
\begin{eqnarray}
f(\omega) S(\omega) & = & \int d^3x
       \delta\langle F^\dagger(x) n(x,\omega)\rangle \nonumber \\
  & = & \frac{1}{\hbar} \int d^3x \int d^3x' F^\dagger(\pol x)
      D^R({\pol x},{\pol x'};\omega) F(\pol x') f(\omega)  .
       \label{eq:3.15}
\end{eqnarray}
Combining eqs.~(\ref{eq:3.13}) and (\ref{eq:3.15}) the imaginary
part of the structure function then gives the total transition
probability associated with $F(\pol x)$.
\begin{equation}
{\it IM} \left[S(\omega)\right] = - \frac{\pi}{\hbar}
    \sum_n \left| \int d^3x'
 \langle\psi_n|{\widetilde n}_x(\pol x')|\psi_0\rangle
  F(\pol x')\right|^2 \delta\left(\omega -
   \frac{E_n-E_0}{\hbar}\right) , ~~~~~E_n\geq E_0 .\label{eq:3.16}
\end{equation}
Note that this quantity is purely negative, where this feature can
be used as a measure of the convergence of the solution.

At this point instead of using the standard route of letting
$f(t)\rightarrow 0$ to recover the linear response equations, we
choose an alternate technique to calculate the response.  In this
case we evolve the system in time and then fourier transform the
result, where $H_{ex}(t)$ is a perturbing function.  In
this case $f(t)$ is chosen to be a Gaussian of the following
form.
\begin{eqnarray}
f(t) & = & \varepsilon e^{-\frac{\alpha}{2} \left(t-{t_\alpha}\right)^2},
            ~~~~~t \geq t_0  \nonumber \\
f(\omega) & = & \varepsilon \sqrt{\frac{2\pi}{\alpha}}
           e^{-\frac{\omega^2}{2\alpha}+\imath\omega t_\alpha},
           \label{eq:3.17}
\end{eqnarray}
where $\varepsilon$ is some small number $(\sim 10^{-5})$, chosen
such that we are in the linear regime.  The parameter, $\alpha$,
is set to be $\sim 1.0~{\rm c/fm}$, which allows for a reasonable
perturbation of collective energies up to $\approx 150$~MeV.
The offset in time, $t_\alpha$ is chosen such that the full
gaussian is approximately within the time regime, $t \geq t_0$.

In practice to evaluate the TDHF equations the time-evolution
operator is used to evolve the system.
\begin{equation}
U(t,t_0) = T \left[ e^{-\frac{\imath}{\hbar} \int_{t_0}^{t}
  dt' {\widehat H}_{tot}(t')}\right] , \label{eq:3.18}
\end{equation}
where $T\left[ ~...~\right]$ denotes time-ordering.  Using
infinitesimal time increments, the time-evolution operator is
approximated by
\begin{eqnarray}
U(t_{n+1},t_n) & = & e^{-\frac{\imath}{\hbar} \int_{t_n}{t_{n+1}}
                   dt' {\widehat H}_{tot}(t')} \nonumber \\
  & \approx & e^{-\frac{\imath}{\hbar} \Delta t
          {\widehat H}_{tot}(t_n+\frac{\Delta t}{2})} \label{eq:3.19} \\
  & \approx & 1 + \sum_{k=1}^{N}\left[ \frac{\left(-\frac{\imath}{\hbar}
            \Delta t {\widehat H}_{tot} \right)^k}{k!}\right] , \nonumber
\end{eqnarray}
where the quantity ${\widehat H}_{tot}^k$ is evaluated by repeated
operations of ${\widehat H}_{tot}$ upon the wave functions.

The procedure is to then choose a particular form for
$F(\pol x)$, using eq.~(\ref{eq:3.17}) for $f(t)$, and time evolve
the system using eq.~(\ref{eq:3.19}).  The Fourier transform in
time of the result then gives us $f(\omega) S(\omega)$, from
which the linear response structure function of the system is
extracted.  For a more detailed discussion please see Ref.~\cite{lrt}.

\section{Results}

The static mean field calculation is analyzed in detail
in Ref.~\cite{umar4}.  For an example of some results using the
static model in a HF+BCS calculation, please see Ref.~\cite{s44},
where the results for various sulphur isotopes are presented.
The pairing correlations are described by using a constant
gap approximation of 0.2~MeV.

The static and dynamic Hartree-Fock calculations are performed
using a 3-dimensional collocation lattice constructed with
B-splines.  Typically a 7th order B-spline is used in a
$(-10, +10~{\rm fm})^3$ box with $20^3-24^3$ grid points.  The calculation
can be performed with or without assuming time-reversal symmetry.
The collective linear response may involve particle-hole
interactions which are spin-dependent and not time-reversal
symmetric.  Therefore, for the correct collective content to be
included one should not
impose time-reversal symmetry in the linear response calculations
\cite{krewald}.  In the results presented in this paper
time-reversal symmetry is not imposed.  In a comparison it is
found that imposing time-reversal symmetry causes small shifts in
the position of the collective modes on the order of
$\approx 0.3$~MeV.

Calculations of isoscalar octupole; and isoscalar
quadrupole collective modes are presented for $^{16}$O using
several parametrizations of the Skyrme interaction.  Here the
parametrizations known as skm* \cite{skm}
and sgII \cite{sgII} are used for comparisons.
A time step of $\Delta t = 0.4$~fm/c is used for the calculations.
It is found that one can perform the time evolution for over
40000 time steps.  The results shown here typically use 16384
time steps for a maximum time of $\approx 6550$~fm/c.

Reasonable results are obtained if the parameter $\varepsilon$ in
eq.~(\ref{eq:3.17}) is chosen to fall in the range
$2.0\times 10^{-4}\leq\varepsilon\leq2\times 10^{-7}$.  By varying
the value of $\varepsilon$, the amplitude of the time-dependent
density fluctuation then scales proportionally to $\varepsilon$,
thus indicating that we are well within the linear regime of the
theory.

The linear response calculations require well converged initial
static HF solutions.  To test for the convergence of the static HF
calculation the energy fluctuation, which is the variance of
$\widehat H$ as given in eq.~(\ref{eq:var}), is minimized.
This measure of convergence tests directly how close
the wave functions are to being eigenstates of $\widehat H$ and is
independent of the iteration step size.  For $^{16}$O it was found
that static HF solutions with energy fluctuation less than
about $1.0\times 10^{-5}$ provided adequate starting points for
the dynamic calculation, although the smaller the energy
fluctuation the better.

The dynamic calculations involve using eq.~(\ref{eq:3.19}) to
evolve the system.  Since $U(t,t')$ is an unitary operator, the
orthonormality of the system is preserved, therefore it is not
necessary to re-orthogonalize the solutions after every time-step.
The stability of the calculation is checked by testing the
preservation of the norm of each wave function.  The number of
terms in the expansion of the exponent in eq.~(\ref{eq:3.19}) is
determined by requiring the norm to be preserved to a certain
accuracy (typically to $\leq 5.0\times 10^{-10}
  \rightarrow 1.0\times 10^{-8}$).

The time-dependent perturbing part of the Hamiltonian is evaluated
when the exponent in eq.~(\ref{eq:3.17}) is greater than some
small number, $\varepsilon_{cut}$.  Since it is not
difficult to evaluate the action of
the external part of the Hamiltonian on the wave function,
$\varepsilon_{cut}$ is chosen to be very small, ($1.0\times 10^{-10}$).
To allow the fourier transform of $f(t)$ to be evaluated easily,
it is necessary to integrate $t$ from $-\infty$ to $\infty$ and
hence we would like the entire Gaussian of the perturbing
function, $f(t)$, to be included into the time evolution to the
desired accuracy.
The parameter $t_\alpha$ is therefore chosen such that the complete
nonzero contribution of the time-dependent perturbation is
included.
$t_\alpha = \Delta t \left( 2 + \sqrt{\left|\frac{2\log{\varepsilon_{cut}}}
{\alpha\Delta t}\right|}\right)$.

Pairing can be easily included using the BCS \cite{ring} or
Lipkin-Nogami \cite{LN} constant pairing strength prescriptions.
These two methods have
been included into the static Hartree-Fock calculations and
can be easily incorporated into the dynamical calculation.
For studies of $\beta-$decay it will be necessary to include
pairing, thus producing calculations of responses to
quasi-RPA excitation modes.

The computations were performed on massively
parallel INTEL iPSC/860 and PARAGON supercomputers.

For the study of the isoscalar quadrupole moment, the perturbing
function $F({\pol x})$, introduced in eq.~(\ref{eq:3.3}), is
chosen to be the mass quadrupole moment, $Q_{20}=2z^2-(x^2+y^2)$.
It turns out that other even multipole modes are also excited
at the same time ({\it i.e.} $Q_{40}$, $Q_{60}$, ...).
One can therefore study the effect of the coupling between the
different excitation modes.  The same holds true for the odd
multipoles.

In the top panel of Fig. 1 the time-dependent evaluation of
the axial quadrupole moment, $\langle {\widehat Q}_{20}(t)\rangle$
is given for $^{16}$O using the skm* parametrization.
\begin{equation}
\langle {\widehat Q}_{20}(t)\rangle = \langle F(x)\rangle =
\int d^3x \delta\langle{\widehat n}(x,t)\rangle F^\dagger(x),
\label{eq:4.1}
\end{equation}
In this case the box is set to be $(-10,+10~{\rm fm})^3$ in size using
$20^3$ grid points.  The periodic nature of the calculation is
clear in this figure, where the smallest oscillation has a period
of about $65$~fm/c.

Both a discrete Fourier transform [FT] and a continuous FT via
Filon's method are used to calculate the FT of
$\langle{\widehat Q}_{20}(t)\rangle$ to give
$\langle{\widehat Q}_{20}(\omega)\rangle=f(\omega)S_{20}(\omega)$.
The frequency grid, $\Delta\omega$ is given by $2\pi/T_{max}$, so
for a finer mesh one need only to time evolve to a larger maximum
time, $T_{max}$.  The continuous FT is also limited by this
criteria.  The discrete and continuous FT's give essentially
identical spectrums, although the continuous FT recovers
$f(\omega)$ more precisely and appears to give a more reasonable
energy-weighted sum rule [EWSR] result.  One can use the analytic
expression for $f(\omega)$ given in eq.~(\ref{eq:3.17}) or a
numerical result, where the difference in the spectrum is
negligible.

The lower panel shows the resulting response calculation.  This
quantity was derived in eq.~(\ref{eq:3.16}) to be purely negative.
The skm* results reflect this property quite well, where one sharp peak
is seen at about $19.8$~MeV.  The calculation recovers about
$92\%$ of the exact EWSR, indicating excellent convergence.

To test the sensitivity of the dynamic calculation to the box
size, the calculation shown in Fig.~1 is repeated using two larger
boxes.  The result in Fig.~1 is compared in Fig.~2 to
calculations using a box~$=(-11,+11~{\rm fm})^3$ with $22^3$ grid
points in the middle panel and one using a
box~$=(-12,+12~{\rm fm})^3$ with $24^3$ grid points in the lower
panel.  In all three cases the grid spacing is set to be 1~fm.
One note is that since the continuum is represented by discrete
states in a box, the resonances are very sharply peaked.  To best
represent these peaks, $T_{max}$ is adjusted and chosen to give
frequency grid locations as close to the peaks as possible.  In the
top panel it ended up that $N=16384$, where
$T_{max}=N\times\Delta t$, is an ideal choice.  For the other two
panels we found that choosing larger $T_{max}$'s and hence larger
$N$'s gave much better spectrums.  One finds though that the
spectrum changes as the box size is increased and in the middle
panel there are 3 large peaks rather than just one.  This
disturbing result indicates that the spectrum as well as the peak
positions appear  to be significantly affected by the box size.
We also find an effect when static boundary conditions are imposed
instead of periodic ones.

In Fig.~3 a calculation similar to Fig.~2 is shown, where the sgII
Skyrme force parametrization is used.  In this case the box also
has a large effect on the calculated spectrum.  The peak positions
shift as well as the number of peaks.  Also shown in the middle
panel is a calculation which uses $22^3$ grid points but with a
denser grid.  In this case there is also a significant change in
the spectrum.

The dependence observed in Figs.~2 and 3 can be traced to the
representation of the continuum as given by the box.  The lowest
energy continuum states in the box correspond to the states with
the fewest nodes.  These states are of course dependent upon the
size of the box.  By increasing the size of the box, the lowest
energy of these continuum states is reduced and the continuum
is better represented in the calculation.  Efforts are underway
to increase the size of the box, by introducing unequal lattice
spacing into the calculation.

In Fig.4 the isoscalar octupole response is shown for the skm* and
sgII effective interactions.  There are 3 sharp resonance
structures at about $6-7$~MeV, $13-14$~MeV and $18$~MeV, where the
sgII resonances are at the slightly higher energies.  These three
peaks are also found in a spherical RPA calculation using the same
effective interactions \cite{pg}.  The spherical calculation finds
an additional broad peak for skm* and sgII centered at an energy
of about $27-28$~MeV.  This peak is weaker than the 3 peaks at
lower energies and is not observed in the three-dimensional linear
response calculation.  In Fig.~4 one can see that the response is
approximately purely negative, but that there are some violations
of this feature.

\section{Summary and Conclusions}

A method for evaluating the linear response theory using TDHF
is formally developed and implemented.  This method allows one to
construct the dynamic calculation directly on top of the static
Hartree-Fock calculation.  Therefore, by performing a sophisticated
and accurate three dimensional static Hartree-Fock calculation, we
also have a corresponding consistent dynamic calculation.  A
coherent description of static ground state properties, such as
binding energies and deformations, is given along with a
description of the collective modes of nuclei.  This feature is
important in studies of $\beta-$decay, where calculated reaction
rates are very sensitive to the deformation properties of the
nucleus.

A collocation basis spline lattice representation is used, which
allows for a much more accurate representation of the gradient
operator and hence a correspondingly accurate overall calculation.
Pairing correlations can be easily included into both the static and
dynamic calculation, thus enabling us to look at quasi-RPA
collective effects.

Example calculations of $^{16}$O are presented for the response
functions corresponding to various isoscalar and isovector
multipole moments.  It is found that the time-evolution of the
system in this full three dimensional calculation contains some
dependences upon the box size, which need to be addressed.

\vfill

\begin{center}
{\bf Acknowledgements}
\end{center}
This work was performed in part under the auspices of the
U.~S. Department of Energy under contracts No.
DE-AC05-84OR21400 with Martin Marietta Energy Systems, Inc. and
DE-FG05-87ER40376 with Vanderbilt University.  This research
has been supported in part by the U.S. Department of Energy,
Office of Scientific Computing under the High Performance
Computing and Communications Program (HPCC) as a Grand
Challenge titled ``the Quantum Structure of Matter''.



\vfill

\begin{center}
{\bf Figure Captions}
\end{center}

\begin{itemize}

\item[Fig.~1]
        {The fluctuations in the isoscalar axial quadrupole moment
        as a function of time are shown in the upper panel for
        $^{16}$O using the skm*
        effective interaction for 16384 time steps.  The size
        of the time step is $0.4$~fm/c with a spatial grid of
        $20^3$ in a cartesian box dimensioned
        $(-10, +10~{\rm fm})^3$.  The response for this
        calculation is shown in the lower panel.  $S(\omega)$ is
        obtained by fourier transforming the upper panel result
        and then dividing by $f(\omega)$.  }

\item[Fig.~2]
        { The lower panel of Fig.~1 is repeated here in the upper
        panel.  The middle and lower panels correspond to the same
        calculation as shown Fig.~1 except the box sizes are
        $(-11, +11~{\rm fm})^3$ and $(-12, +12~{\rm fm})^3$,
        respectively and the grid lattices are $22^3$ and $24^3$,
        respectively.  The number of time steps used, $N$ are labeled
        on the graphs.  }

\item[Fig.~3]
       {The same as Fig.~2, except the sgII parametrization of the
        Skyrme force is used and the number of time steps in the
        time evolution is different.  Also shown in the middle
        panel as the dashed curve is a calculation using a
        $22^3$ lattice with a box size of
        $(-10.6, +10.6~{\rm fm})^3$. }

\item[Fig.~4]
       {The imaginary part of the response function corresponding
        to the isoscalar octupole moment is shown for $^{16}$O
        and executed for 16384 time steps.  The skm* and sgII
        parametrizations of the Skyrme force are used in the
        upper and lower panels, repectively with varying lattice
        grid and box sizes as labelled in the figure. }

\end{itemize}


\begin{thebibliography}{99}

\bibitem{li11} C. R. Chinn, J. Decharg\'e and J.-F. Berger,
              `Correlations in a Many-Body Calculation of $^{11}$Li',
              to be submitted to Phys. Rev. {\bf C}.

\bibitem{bf} J. Blocki and H. Flocard, Phys. Lett. {\bf 85B},
             163 (1979).

\bibitem{ASU}
          A. S. Umar, M. R. Strayer, P. -G. Reinhard,
          K. T. R. Davies, and S. -J. Lee, Phys. Rev. C {\bf 40}, 706
          (1989).
\bibitem{DaKo}
          K. T. R. Davies and S. E. Koonin, Phys. Rev. C
          {\bf 23}, 2042 (1981).
\bibitem{HN}
          P. Hoodbhoy and J. W. Negele, Nucl. Phys.
          {\bf A288}, 23 (1977).
\bibitem{KDM}
          S. E. Koonin, K. T. R. Davies, V. Maruhn-Rezwani,
          H. Feldmeier, S. J. Krieger, and J. W. Negele, Phys. Rev. C
          {\bf 15}, 1359 (1977).

\bibitem{Boor}
           C. De Boor, {\it Practical Guide to Splines},
           (Springer-Verlag, New York, 1978).

\bibitem{bottcher} C. Bottcher and M. R. Strayer, Ann. of Phys.
                   {\bf 175}, 64 (1987).

\bibitem{umar3} A. S. Umar, J. Wu, M. R. Strayer and C.~Bottcher,
                J. of Comp. Phys. {\bf 93}, 426 (1991).

\bibitem{BSU}
           C. Bottcher, M. R. Strayer, A. S. Umar, and
           P. -G. Reinhard, Phys. Rev. A {\bf 40}, 4182 (1989).

\bibitem{umar1} A. S. Umar, M. R. Strayer, R. Y. Cusson, P.-G.~Reinhard,
                and D.~A.~Bromley, Phys. Rev. {\bf C 32}, 172 (1985).

\bibitem{fetter} A. L. Fetter and J. D. Walecka, {\underline
                 {Quantum Theory of Many-Particle Systems}},
                 (St. Louis: McGraw-Hill Book Co., 1971).

\bibitem{lrt} C. R. Chinn, A. S. Umar, and M. R. Strayer,
              `Time-Dependent Evaluation of Linear Response Theory',
              to be submitted to Phys. Rev. {\bf C}.

\bibitem{umar4} A. S. Umar, M. R. Strayer, J.-S. Wu, D.~J.~Dean and
                M.~C.~G\"{u}\c{c}l\"{u}, Phys. Rev. {\bf C 44}, 2512 (1991).

\bibitem{s44} T.R. Werner, J.A. Sheikh, W. Nazarewicz, M.R. Strayer,
              A.S. Umar and M. Misu, ``Shape Coexistence Around
              $^{44}_{16}S_{28}$:  The Deformed $N=28$ Region'',
              submitted to Phys. Lett.

\bibitem{krewald} S. Krewald, V. Klemt, J. Speth and A. Faessler,
                  Nucl. Phys. {\bf A281}, 166 (1977).

\bibitem{skm} J. Bartelm, O. Quentin, M. Brack, C. Guet and H. B.
              Hakansson, Nucl. Phys. {\bf A 386}, 79 (1982).

\bibitem{sgII} N. Van Giai and H. Sagawa, Nucl. Phys. {\bf A371},
               1 (1981).

\bibitem{ring} P. Ring and P. Schuck, {\underline {The Nuclear
               Many-Body Problem}}, (New York: Springer-Verlag 1980).

\bibitem{LN} H. C. Pradhan, Y. Nogami and J. Law, Nucl. Phys.
             {\bf A201}, 357 (1973); Y. Nogami, Phys. Rev.
             {\bf 134}, B313 (1964).

\bibitem{pg} Y. Gambhir, P.-G. Reinhard, Ann Phys. (Leipzig) {\bf 1}
             (1992), page 598; P.-G. Reinhard, Ann. Phys. (Leipzig)
             {\bf 1} (1992), p 632; P.-G. Reinhard, private
             communication.

\bibitem{data} K. T. Kn\"opfle, G. J. Wagner, H.~Breuer, M.~Rogge,
               C.~Mayer-B\"oricke, Phys. Rev. Lett. {\bf 35}, 779 (1975).

\end{thebibliography}
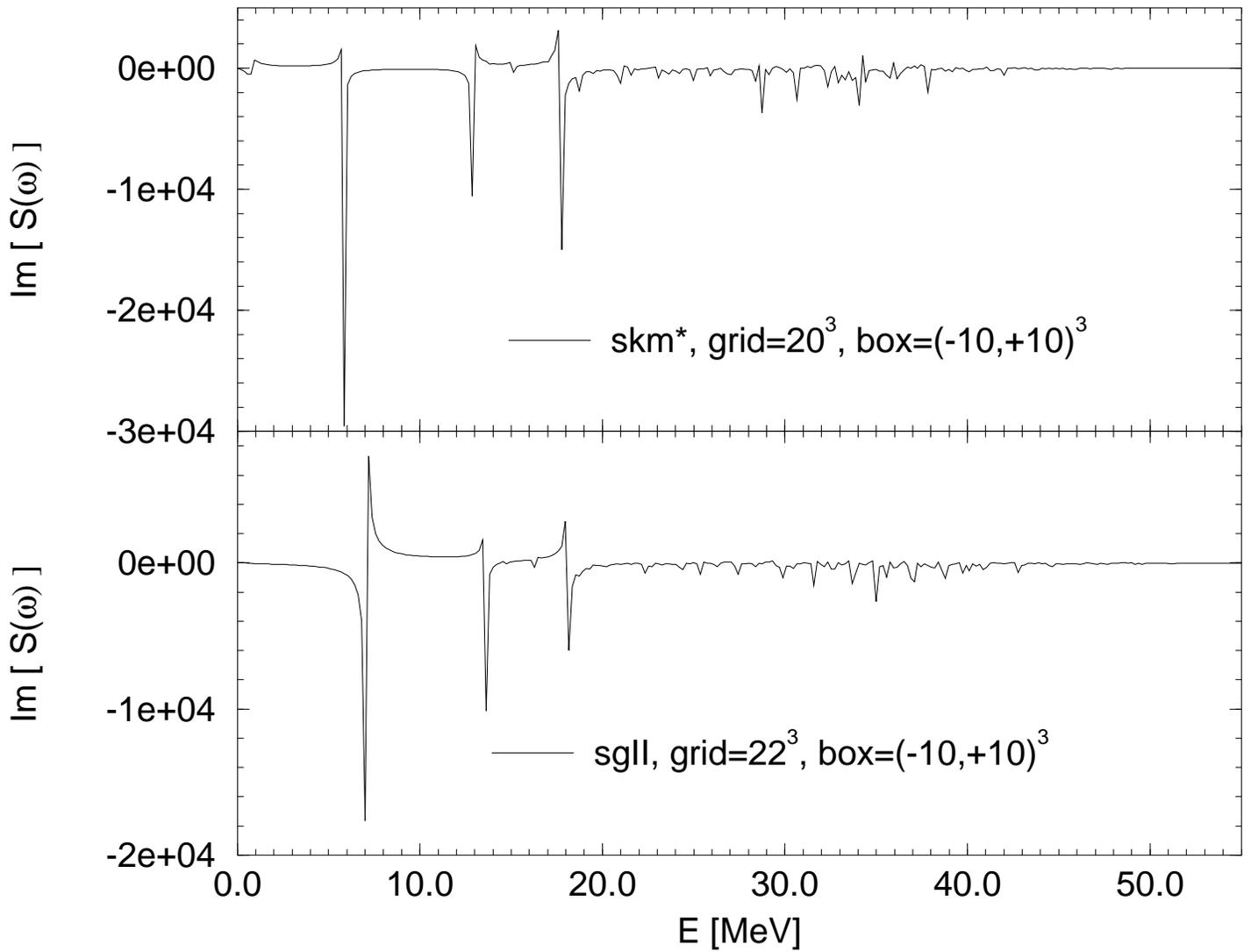
\end{document}